# SUDDEN INTENSITY INCREASES AND RADIAL GRADIENT CHANGES OF COSMIC RAY MeV ELECTRONS AND PROTONS OBSERVED AT VOYAGER 1 BEYOND 111 AU IN THE HELIOSHEATH

**W.R Webber[1], F.B. McDonald[2], A.C. Cummings[3], E.C. Stone[3], B. Heikkila[5] and N. Lal[5]**

1. New Mexico State University, Department of Astronomy, Las Cruces, NM, 88003, USA
2. University of Maryland, Institute of Physical Science and Technology, College Park, MD, 20742, USA
3. California Institute of Technology, Downs Laboratory, Pasadena, CA 91125, USA
4. NASA/Goddard Space Flight Center, Greenbelt, MD 20771, USA

## ABSTRACT

Voyager 1 has entered regions of different propagation conditions for energetic cosmic rays in the outer heliosheath beginning at a distance of about 111 AU from the Sun. This conclusion is based on the fact that the low energy 6-14 MeV galactic electron intensity suddenly increased by ~20% over a time period ≤ 10 days and the electron radial intensity gradient abruptly decreased from ~19%/AU to ~8%/AU at 2009.7 at a radial distance of 111.2 AU. A sudden radial gradient change was also observed at this time for >200 MeV protons. The gradients were constant during the time period before and after the electron increase. At about 2011.2 at a distance of 116.6 AU a second abrupt intensity increase was observed, this time for both electrons and protons. The increase for electrons was ~25% and occurred over a time period ~15 days or less. For >200 MeV protons the increase at this time was ~5% (unusually large) and occurred over a longer time period ~50 days. Between about 2011.2 and 2011.6, radial intensity gradients ~18%/AU and 3%/AU were observed for electrons and protons, respectively. These gradients were very similar to those observed for these particles before the 1[st] sudden increase at 2009.7. These large positive gradients observed after 2011.2 indicate that V1, although it has entered a different propagation region, is still within the overall heliospheric modulating region at least up to a time ~2011.6 (118.0 AU). In this paper we will discuss these events in more detail and consider possibilities for their explanation that have recently been suggested.

**Introduction**

The study of low energy cosmic ray electrons (≤100 MeV) has historically been an enigma. From the modulation point of view these electrons with their low rigidity, high speed and relatively steep energy spectrum behave quite differently than higher energy electrons and nuclei. In particular they respond directly to the diffusion coefficient that determines their propagation or modulation. Also these electrons are the low energy tail of the galactic cosmic ray electron spectrum whose intensity at higher energies can be estimated from the galactic radio-synchrotron intensity and spectrum. Comparisons of the measured electron intensities near the Earth in this low energy range with those estimated to exist in the nearby galaxy that are required to produce the radio synchrotron emission suggest an intensity difference of a factor ~$10^2$-$10^3$ that must be accounted for by solar modulation, a much larger modulation factor than required for higher rigidity electrons.

The CRS electron experiment on Voyagers 1 and 2 measures electrons in the energy range 6-120 MeV and is well suited for the study of these electrons and how the intensities measured near the Earth evolve into those estimated to exist in nearby interstellar space (Stone, et al., 1977).

This involves measuring the electron intensity at V1 and V2 as these spacecraft move outward through the heliosphere. At the outer boundary of the heliosheath the intensity is usually assumed to be equal to that in nearby interstellar space. So the important parameters to be determined in this study are: 1) The electron intensity near the location of the heliospheric termination shock (HTS) which marks the beginning of the heliosheath, and 2) The electron intensity near the location of the outer heliosheath boundary.

The intensity at the HTS and the HTS distance itself have already been determined by V1 and V2 as they move outward to their locations of 117.5 and 96 AU respectively at 2011.5. The HTS distance is 94 AU at V1 in the N-hemisphere and 84 AU at V2 in the S-hemisphere (Stone, et al., 2008) and the 6-14 MeV electron intensity at these locations is roughly (within a factor ~2) the same as measured at the Earth (McDonald, et al., 2007, 2009). So, most of the increase in the electron intensity necessary to reach the estimated interstellar value beyond the heliosheath appears to occur in the heliosheath region.

Indeed V1 has measured large radial intensity gradients of electrons beyond the HTS (McDonald, et al., 2009; Caballero-Lopez, et al., 2009; Nkosi, et al., 2011). The electron gradient that is determined for V1 beyond the HTS has been very irregular, because of transients moving outward through the heliosheath (Webber, et al., 2009; Washimi, et al., 2011). This has made it difficult to determine the electron radial gradients in the inner heliosheath.

After about 2008.5, when V1 was at least 12 AU beyond the distance of its initial crossing of the HTS and at a time when the last outward major interplanetary transients from solar cycle #23 had propagated beyond the modulation region, the electron intensity started a smooth regular increase which continued for over a year, corresponding to ~4 AU of outward movement. At 2009.7 a sudden increase of electron intensity by ~20% occurred, followed by another time period of smooth regular increase but with a much smaller radial intensity gradient continuing to 2011.2. At 2011.2 a second, even larger sudden increase occurred in the 6-14 electron intensity followed by a resumption of a large radial gradient similar to that seen before 2009.7.

For cosmic ray protons >200 MeV these increases are also observed but are more gradual. A sudden change in the radial intensity gradient of protons is observed at 2009.7 and again at 2011.2, however. It is this sequence of changes, possibly indicating the passage of V1 into regions of different propagation in the outer heliosheath, that we wish to discuss in this paper.

**The Data**

The 5-period running average of the daily intensities of 6-14 MeV electrons and >200 MeV protons at V1 from the time that V1 crossed the HTS in late 2004 to 2011.5 is shown in Figure 1. It is seen that the V1 electron intensity increases, more or less continuously from a low value, which has a significant background at the time of the HTS crossing at the end of 2004, to a value in 2011 that is >50 times the initial intensity. The >200 MeV proton intensity also increases more or less continuously with a much smaller overall increase, ~ a factor of 1.8 from the time of the HTS crossing to mid 2011.

At V1, for the first 3.5 years after the shock crossing, it is difficult to determine a unique average radial gradient for electrons from the V1 data itself because of the transient intensity variations. Some of these temporal electron variations are time coincident with transient decreases of >200 MeV protons (Webber, et al., 2009). These transient decreases could be caused by the largest IP shocks as they propagate beyond the HTS, through the heliosheath, ultimately reaching the HP (Webber, et al., 2007, 2009; Washimi, et al., 2011). A simple average of the 6-14 MeV electron radial gradient between 2005.5-2008.5, during a time in which the intensity increased by a factor of ~5, would lead to an average radial gradient $G_r$, defined as $G_r = (r_2-r_1)^{-1} \cdot \ell n \ (J_2/J_1)$ where $J_2/J_1$ are the intensities at radii $r_2, r_1$; of ~53%/year or about 15%/AU (see McDonald, et al., 2007, 2009).

After about 2008.5 the rate of increase in 6-14 MeV electron intensity at V1 becomes smoother and more regular. This later time period is shown in expanded form in Figure 2.

From about 2008.5 – 2009.7 (4.3 AU in distance) the total increase of 6-14 MeV electrons at V1 is 2.29 ±0.18 which corresponds to an average radial gradient of 19.2 ±1.5%/AU if the increase is all due to radial effects. At 2009.7 (111.2 AU) the electron intensity suddenly increases by ~20% in a time period ≤10 days after which it resumes a smooth regular increase but with a different radial intensity gradient.

Over the time interval from 2008.5-2009.7 the increase of >200 MeV protons is 1.125±0.01 which is equivalent to a radial gradient of 2.7 ± 0.25%/AU.

The jump in electron intensity at 2009.7 is seen in other electron channels in the energy range of 2.5-120 MeV covered by the V1 instrument, most notably the 14-26 MeV channel. Here we concentrate on the 6-14 MeV channel because of its better statistics (2-3% variations on a day to day basis).

Between 2009.7 and 2011.2 the electron intensity increase is a factor of 1.55 ± 0.05 corresponding to an average radial gradient of 8.1 ± 0.7%/AU. For the >200 MeV protons these values are 1.043 ± 0.005 and 0.97 ± 0.10%/AU, respectively.

So the average radial gradients that are measured for both electrons and protons between 2009.7 and 2011.2 are only about 0.4 of those in the earlier time period from 2008.5 to 2009.7. We believe that these features indicate that V1 has crossed into a region in the outer heliosheath in which the propagation conditions have changed significantly at 2009.7. This is discussed in more detail in the following section.

At 2011.2 (116 AU) there is a second sudden increase in electron intensity, ~25% over a time period ~15 days and a corresponding increase of ~5% in the proton intensity over at time period ~50 days.  This is the largest change in >200 MeV proton intensity observed since the HTS crossing over 6 years earlier.  We therefore believe that V1 has crossed into still another new propagation region at about 2011.2.

Between 2011.3 and 2011.6 the electron radial gradient becomes ~18 ±3%/AU and the proton gradient becomes ~3.5 ±0.5%/AU.  These gradients are similar to those before 2009.7 when V1 was inside 111 AU.

## Discussion of the Electron and Proton Radial Gradients after 2008.5 and the Sudden Intensity Changes of Electrons at 2009.7 and 2011.2

Conditions in the outer heliosheath are not well known.  However, a significant step in understanding this regime has been made recently by Pogorelev, et al., 2009; see also Washimi, et al., 2011, and also Borovikov, et al, 2011, and Florinski, 2011.  These authors have modeled a heliospheric current sheet (HCS) of constant tilt as it propagates in the supersonic solar wind upstream of the HTS and then beyond the HTS into the heliosheath.  These calculations show that the latitude extent of the HCS rapidly increases as the boundary between the solar wind and the interstellar medium is approached and the solar wind flow diverts to carry the plasma down the heliotail.  As a result of this increasing latitude extent, V1 which is at ~35° N latitude, well above the maximum N extent of the HCS near the Sun reported by the Wilcox Solar Observatory (http://wso.stanford.edu), would be expected to enter a new region of solar wind flow in the outer heliosheath prior to reaching the heliopause (HP).  According to Florinski, 2011, particle transport across this region would be easier, thus accounting for the observed smaller gradients.

Also using a computer intensive 3D MHD model of the heliosphere with emphasis on the heliosheath, Washimi et al., 2010, have identified at least one possible major structure in the outer heliosheath that lies between about 8-18 AU inside of the HP.  This structure is called the "magnetic wall".  It appears to have characteristics that could provide a sudden jump in electron intensity and a change in radial intensity gradients inside and outside of the wall as the effective diffusion coefficient changes.

Recent magnetic field data from Voyager 1 covering this time period (Burlaga and Ness, 2010), shows a large spike in the total magnetic field intensity at exactly 2009.70.  This spike has a peak magnitude ~0.3 nT, 3-5 times the average value of B and the spike has duration ~10 days.  It is also near the onset of a time period when the field direction suddenly changes by 180° from 90° (negative) to 270° (positive). The duration of the 270° field lasts for over 100 days. These are the most prominent features in the magnetic field magnitude and direction observed at V1 since it crossed the HTS almost 5 years earlier.

Recent observations of the outward flowing solar plasma by Krimigis, et al., 2011, indicate that the radial speed of this plasma approaches zero in late 2009 and has remained near or less than zero until the present time.

In this connection the second sudden electron intensity increase in the heliosheath at 2011.2 at V1 may be even more significant than the first. The second sudden intensity increase of both electrons and >200 MeV/protons at 2011.2 (average rigidity ~1.5 GV for protons) is unprecedented in magnitude in the heliosheath and even exceeds the first increase occurring ~1.5 years earlier. At this time the radial gradients of both electrons and protons also changed, reverting to the values ~18%/AU and 3%/AU that were observed earlier between ~2008.5 and 2009.7.

## Summary and Conclusions

After crossing the HTS in late 2004, the 6-14 MeV (galactic) electron intensity measured at V1 increased rapidly and irregularly. By about 2008.5 this intensity was ~5 times that measured when V1 crossed the shock.

After 2008.5 the electron intensity at V1 continued to increase rapidly but now much more smoothly so that a more accurate local radial intensity gradient could be determined. This average gradient was determined to be ~18.5±1.5%/AU from 2008.5 to 2009.7 (111.2 AU) at which time the electron intensity suddenly increased by ~20% over a period ≤10 days. After that the intensity continued its smooth increase, but now corresponding to a much smaller radial gradient of 8 ± 1.0%/AU which continued until 2011.2 when V1 was at 116.6 AU. At about 2009.7 the >200 MeV proton gradient also changed suddenly from ~2.7 ±0.25%/AU between 2008.5 and 2009.7 to 0.97 ± 0.10%/AU after 2009.7. For both electrons and protons the radial gradient after 2009.7 was only ~0.4 of that measured earlier indicating a significant change in propagation conditions at that time.

At about 2011.2 there was a second sudden increase in both the electron and proton intensities. The electrons increased by ~25% over a period of 10-15 days whereas the protons increased by ~5% over a longer period, 30-50 days. After this intensity jump the radial intensity gradients of both electrons and protons increased by a factor ~2.5, reverting to their original values prior to 2009.7. The size of this intensity change, particularly for the high rigidity protons as well as the gradient changes for both electrons and protons indicates a second significant change in propagation conditions occurred at this location.

We believe that these sudden changes in intensity and in the gradients of electrons and protons are evidence that V1 crossed into regions of significantly different propagation conditions at 2009.7 and again at about 2011.2. At the time of the first crossing the radial gradient of both electrons and nuclei decreased by a factor ~2.5. This could be explained, for example, if the effective radial diffusion coefficient increased by a factor ~2.5 at that time.

The two boundaries at 2009.7 and 2011.2 could part of structure in the outer heliosheath and perhaps compatible with the recent observations of the solar wind plasma which show that the radial solar wind speed approached zero at about 2009.5 (Krimigis, et al., 2011). It could also be compatible with descriptions of the radial extent of the HCS fold back, accompanied by a plasma flow directed towards the tail of the heliosphere just inside the HP as described by Pogorelov, et al., 2009, Borovikov, et al., 2011 and Florinski, 2011. If this is the case, V1 has been passing through regions of plasma flow directed towards the heliotail beyond 111 AU. It is unlikely that V1 has reached the undisturbed local interstellar medium at 2011.6 ≡118 AU, however, because of the significant positive radial intensity gradients that are still observed for both electron and protons after the sudden increase at 2011.2.

More data including magnetic field data, which has been useful in helping to understand the origin of the first sudden cosmic ray intensity and gradient changes at 2009.7, will be needed to help understand the full implications of these sudden cosmic ray intensity and radial gradient changes in the outermost heliosheath and what they imply for the proximity of the heliopause or the entry of V1 into a fully interstellar dominated medium where the radial intensity gradient might be expected to approach zero.

Acknowledgements: The authors all appreciate the support of the Voyager program by JPL. The data used here comes from both the Voyager CRS experiments web-site (http://voyager.gsfc.nasa.gov) and from internally generated CRS documents.

# REFERENCES

Borovikov, S.N., N.V. Pogorelov, L.F. Burlaga and J.D. Richardson, (2011), Plasma near the heliosheath: Observations and modeling, Ap.J. Lett. 728, 1-5

Burlaga, L.F. and N.F. Ness (2010), Sectors and large scale magnetic field strength fluctuations in the heliosheath near 110 AU: Voyager 1, 2009, Ap.J., 735, 1306-1316

Caballero-Lopez, R.A., H. Moraal and F.B. McDonald, (2010), The modulation of galactic cosmic ray electrons in the heliosheath, Ap.J., 725, 121-127

Florinski, V., (2011), The transport of cosmic rays in the distant heliosheath, Advances in Space Res., 48, 308-313

Krimigis, S.M., E.C. Roelof, R.B. Decker and M.E. Hill, 2011, Zero outward flow velocity for plasma in the heliosheath transition layers, Nature, 474, 359-361, doi:10.1038/Nature 10115

McDonald, F.B., et al., (2007), Voyager observations of galactic cosmic ray electrons in the heliosheath, AGU, Fall Meeting, abstract #SH11A-08

McDonald, F.B., et al., (2009), Voyager observations of galactic ions and electrons in the heliosheath, Proceedings of the 31$^{st}$ ICRC, Lodz

Nkosi, G.S., M.S. Potgieter and W.R. Webber, (2011), Modeling of low energy electrons in the heliosheath, J. Adv. Space Res., doi:10.1016/jasr2011.06.017

Pogorelov, N.V., S.N. Borovikov, G.P. Zank and T. Ogino, (2009), Three dimensional features of the outer heliosphere due to coupling between the interstellar and interplanetary fields; The wavy heliospheric current sheet, Ap.J., 696, 1478-1490

Stone, E.C., et al., (1977), Cosmic ray investigation for the Voyager missions: Energetic particle studies in the outer heliosphere – and beyond, Space Sci. Rev., 21, 355-376 doi: 10.1007/BF00211546

Stone, E.C., et al., (2008), An asymmetric solar wind termination shock, Nature, 454, 71-74, doi: 10.1038/nature07022

Washimi, H., G.P. Zank, Q. Hu and T. Tanaka, (2011), Realistic and time-varying outer heliospheric modeling by three-dimensional MHD simulation, Mon. Not. R. Astron. Soc., 416, 1475-1485

Webber, W.R., A.C. Cummings, F.B. McDonald, E.C. Stone, B. Heikkila and N. Lal, (2007), Passage of a large interplanetary shock from the inner heliosphere to the heliospheric termination shock and beyond: Its effects on cosmic rays at V1 and V2, Geophys. Res. Lett., 34, L20107, doi:10.1029/2007GL31339

Webber, W.R., A.C. Cummings, F.B. McDonald, E.C. Stone, B. Heikkila and N. Lal, (2009), Transient intensity changes of cosmic rays beyond the heliospheric termination shock as observed at Voyager 1, JGR, 114, A07108, doi:10.1029/2009JA014156

**FIGURE CAPTIONS**

**Figure 1:**  The 5-day running averages of the 6-14 MeV electron and >200 MeV proton intensities measured at V1 from 2004.0 to the present time.  V1 crossed the HTS at 2004.96 at a distance of 94.0 AU.  The >200 MeV proton intensities are shown on the right hand axis on a greatly expanded linear scale.

**Figure 2:**  The 5-day running average of the 6-14 MeV electrons and >200 MeV protons intensities measured at V1 from 2008.5 to the present time.  The first sudden intensity increase of electrons at 2009.70 (111.2 AU) and the change in radial gradients of both electrons and protons before and after the 1$^{st}$ increase are clearly evident.  So are the second sudden intensity increases of both electrons and protons at 2011.2 (116 AU).  The continuing increase of electron and proton intensities after 2011.3 is also a notable feature.  The >200 MeV proton intensities are shown on the right hand axis on a greatly expanded linear scale.

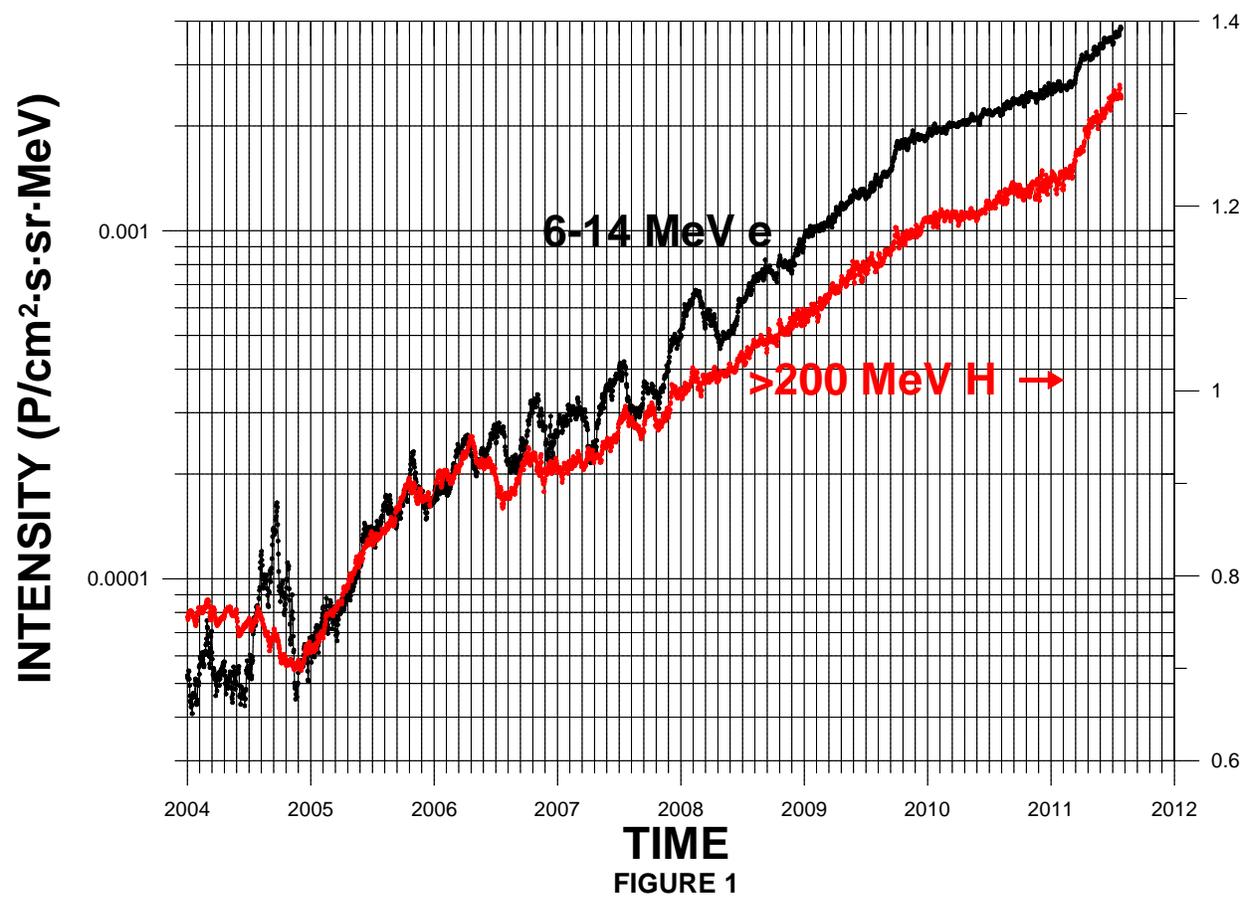

**FIGURE 1**

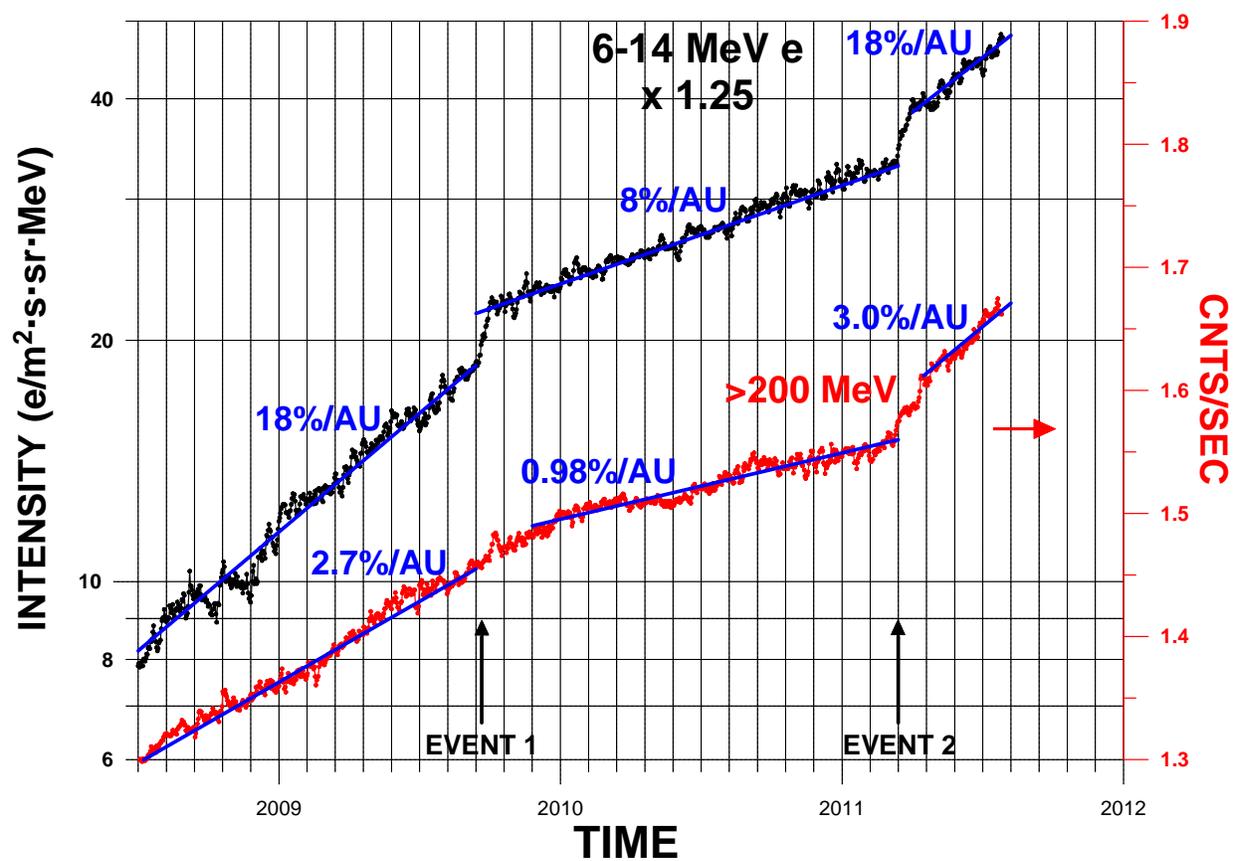

**FIGURE 2**